\documentstyle[twocolumn,seceq,epsf]{jpsj}

\def\runtitle{Magnetization plateaus in polymerized chains
}
\def\runauthor{W. {\sc Chen}, K. {\sc Hida} and B. C. {\sc Sanctuary}}

\title
{
Magnetization plateaus and phase diagram in polymerized $S=1/2$ XXZ chains
}

\author
{
Wei {\sc Chen}\footnote{E-mail: chen@chemistry.mcgill.ca}, Kazuo {\sc Hida}$^1$\footnote{E-mail: hida@phy.saitama-u.ac.jp} and B. C. {\sc Sanctuary}
}

\inst
{
Department of Chemistry, McGill University, Montreal, PQ, Canada H3A 2K6 \\
$^1$Department of Physics, Saitama University, Urawa, Saitama, Japan 338-8570\\
}

\recdate
{
\today
}

\abst
{The magnetization plateaus of $p$-merized  $S=1/2$ XXZ chains are studied for general values of $p$. Two plateau-non-plateau critical lines and one plateau-plateau critical line are found for each value of $p$. The universality class of the plateau-non-plateau transition  belongs to Brezinskii-Kosterlitz-Thouless (BKT) type and that of the plateau-plateau transition, to the Gaussian type. The critical points are determined by level spectroscopic analysis of the numerical diagonalization results for $4 \le p \le 8$. The multicritical points are calculated using the integral equations based on the Bethe ansatz solution of  the XXZ model. The behavior of multicritical points are analyzed in detail for large $p$. It is found that the plateau region is enhanced with the increase of periodicity $p$ although the non-plateau region persists as far as $p$ is finite.
}

\kword
{
magnetization plateau, level spectroscopy, exact diagonalization, BKT transition, twisted boundary condition
}

\begin{document}
\sloppy
\maketitle
\section{ Introduction}
One-dimensional quantum spin systems are attracting both experimental and theoretical interest\cite{rev}. Various phenomena have been discovered which result from strong spin-spin correlations and strong quantum fluctuations. Among these phenomena, the magnetization plateaus are attracting broad interest as an essentially macroscopic quantum phenomenon in which macroscopic magnetization is quantized to fractional values of the saturated magnetization value. Such plateaus are predicted and/or observed in many low dimensional spin systems\cite{mo,hk,okamoto,tone,to2,to,ts2,na,ak,kiok,kiok1,chn,chs,naru,WS}. The plateau state can be also regarded as a spin gap state with non-zero magnetization. Oshikawa, Yamanaka and Affleck\cite{mo} proposed the  necessary condition for the magnetization plateaus as $p(S-m^{z})=q\equiv \mbox{integer}$, where $p$ is the periodicity of the magnetic ground state in the thermodynamic limit, $S$ is the magnitude of the spin and $m^{z}$ is the magnetization per site. 

As seen from the above criterion, the magnetization plateau appears due to the periodic superstructure of the ground state. Therefore it is of interest to investigate how the plateau develops with the periodicity $p$ for a specific model. In the present work, we investigate this problem concentrating on the highest plateau ($q=1$) of one dimensional $p$-merized $S=1/2$ XXZ model  for general values of $p$. 

This paper is organized as follows. In the next section, the model Hamiltonian is defined and the numerical results of the phase diagram is presented. For each value of $p$, we find one Gaussian plateau-plateau critical line and two plateau-non-plateau  Brezinskii-Kosterlitz-Thouless (BKT)  critical lines. Three critical lines merge at a multicritical point. The $p$-dependence of the multicritical point is investigated in detail by Bethe ansatz method\cite{bh,yy,dc} in {\S} 3. The final section is devoted to a summary and discussion.

\section{Numerical Results}
\subsection{Model Hamiltonian}
The Hamiltonian of $p$-merized $S=1/2$ XXZ chain in the magnetic field is given by
\begin{eqnarray}
\label{model}
{\cal H}&=&\sum_{l=1}^{L}\big \{ \sum_{i=0}^{p-2} (1-t){\cal H}_{pl+i,pl+i+1}(\Delta)\nonumber \\
&+&[1+(p-1)t]{\cal H}_{pl+p-1,pl+p}(\Delta) \Big \} \\
&-&g\mu_{\rm B}H\sum_{l=1}^{pL}S_{l}^{z},\nonumber
\end{eqnarray}
where 
\begin{equation}
{\cal H}_{l,l+1}(\Delta)=S_{l}^{x}S_{l+1}^{x}+S_{l}^{y}S_{l+1}^{y}+\Delta S_{l}^{z}S_{l+1}^{z},\ \ -1 \leq t \leq 1
\end{equation}
The magnitude of polymerization, periodicity, anisotropy parameter, magnetic field, the electronic $g$-factor and  Bohr magneton are represented by $t$, $p$, $\Delta$, $H, g$ and $\mu_{\rm B}$, respectively. In the following, we take the unit $g\mu_{\rm B}=1$. 

For $t=0$, the low energy sector of this model with arbitrary magnetization can be expressed by a Gaussian model using the Jordan-Wigner transformation and bosonization technique as,
\begin{equation}
\label{eq:G}
{\cal H_{\mbox{G}}}= \frac{1}{2\pi} \int {dx \Big[ v_{\mbox{s}}K (\pi\Pi)^{2}+\frac{v_{\mbox{s}}}{K}(\frac{\partial \phi}{\partial x})^{2} \Big]},
\end{equation}
where $\phi$ is the boson operator restricted to the range $0 \leq \phi < \sqrt{2}\pi$ and $\Pi$ is the momentum density conjugate to $\phi$ which satisfies $[\phi(x),\Pi(x')]=i\delta(x-x')$. The Luttinger liquid parameter and velocity of a spin wave are represented by $K$ and $v_{\mbox{s}}$, respectively. This model can be described by the conformal field theory with conformal charge $c=1$.

If $t$ is finite but small, the low energy sector of our model (\ref{model}) with $m^z=1/2-1/p$ reduces to the sine-Gordon model\cite{kiok}, which is given by
\begin{eqnarray}
\label{eq:sg}
{\cal H_{\mbox{SG}}}&=& \frac{1}{2\pi} \int {dx \Big[ v_{\mbox{s}}K (\pi\Pi)^{2}+\frac{v_{\mbox{s}}}{K}(\frac{\partial \phi}{\partial x})^{2} \Big]}\nonumber \\
&+& \frac{y_{1}v_{\mbox{s}}}{2\pi a^{2}} \int{dx \cos \sqrt{2}\phi},
\end{eqnarray}
where the $a$ is the lattice constant, the $\cos \sqrt{2} \phi$ term generates the energy gap and its coefficient $y_{1}$  is given by $y_{1}=2\pi t/v_{\mbox{s}} \propto t$.\cite{kiok} For small $t$, $y_1$ is small, so that this term is regarded as a perturbation in calculating the low energy properties.  From the flow diagram of the lowest order  renormalization group equation of the sine-Gordon model\cite{kiok}, we know that there are three critical lines which belong to two different universality classes. Two are the plateau-non-plateau transition lines $K-4=2\mid y_1\mid$. These are BKT transitions. Another is the plateau-plateau transition at $t=0$ for $1\leq K<4$. This is a Gaussian transition. The three critical lines merge  at a multicritical point $(t,K)=(0,4)$.

\subsection{BKT critical point}

It is difficult to estimate precisely the  BKT critical point from standard finite size analysis of the numerical calculation data. Nomura and Kitazawa\cite{no3} proposed to use the level spectroscopic method\cite{nomura} with twisted boundary condition to overcome this difficulty. This method has been successfully applied to the plateau-non-plateau transition of the $S=1/2$ trimerized XXZ chain by Okamoto and Kitazawa.\cite{kiok} We therefore employ this method for the present model.  Here we do not explain the motivation and background of this method, because these are well described in ref. \citen{kiok}.

The finite size critical point is determined from the crossing point of $\Delta E_{0,2}$ and the lower of $\Delta E^{c,s}_{1/2,0}$ defined by,
\begin{eqnarray}
\Delta E_{0,2} &=&\frac{1}{2}\Big\{E_{0}(M_{\rm p}+2,0,1)+E_{0}(M_{\rm p}-2,0,1)\Big\} \nonumber \\
&-&E_{0}(M_{\rm p},0,1).
\end{eqnarray}
and  
\begin{equation}
\Delta E_{1/2,0}^{c}=E^{TBC}(M_{\rm p},1)-E(M_{\rm p},0,1)
\end{equation}
\begin{equation}
\Delta E_{1/2,0}^{s}=E^{TBC}(M_{\rm p},-1)-E(M_{\rm p},0,1)
\end{equation}
where $E_{0}(M^{z},k,P)$ is the lowest energy under periodic boundary condition with magnetization $M^{z} (=Nm^z\equiv \displaystyle\sum_{l=1}^{N} S_{l}^{z} )$, wave number $k$ and parity $P$. The magnetization on the plateau is denoted by $M_{p} (\equiv pL(1/2-1/p)$. The energy  $E^{TBC}(M^{z},P)$ is the lowest energy with the twisted boundary conditions with  magnetization $M^{z}$ and parity $P$. 

\begin{figure}
\epsfxsize=70mm 
\centerline{\epsfbox{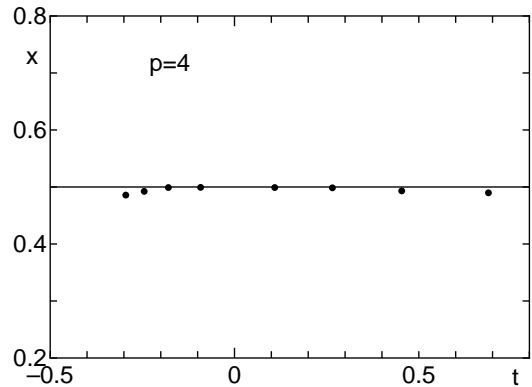}}
\caption{Extrapolated values of the averaged scaling dimension $x$ on the critical points for $p=4$. The solid line is $x=0.5$}
\label{x}
\end{figure}

To confirm the reliability of this method, we have also checked that the following average 
\begin{equation}
\label{ve}
x=\frac{x_{1/2,0}^{c}(L)+3x_{1/2}^{s}(L)}{4}.
\end{equation}
is close to 0.5 at the critical points\cite{kiok} where $x_{1/2,0}^{c,s}(L)$ are the scaling dimensions corresponding to  $E_{1/2,0}^{c,s}$ defined by
\begin{equation}
x_{1/2,0}^{c,s}=\frac{L}{2\pi v_{\rm s}}\Delta E_{1/2,0}^{c,s}.
\end{equation}
Here $v_{s}$ is the spin wave velocity given by
\begin{equation}
v_{\rm s}=\lim_{L\rightarrow \infty} \frac{L}{2\pi}[E_{M_p, k_{1}}(L)-E_{M_p}(L)],
\end{equation}
where $E_{M_p, k_{1}}(L)$ is the energy of the excited state with wave number $k_{1}=\frac{2\pi}{L}$ and $M^{z}=M_{\rm p}$. The results are shown in Fig. \ref{x} for $p=4$ which confirms that $x=0.5$ holds with good accuracy.

\begin{figure}
\epsfxsize=70mm 
\centerline{\epsfbox{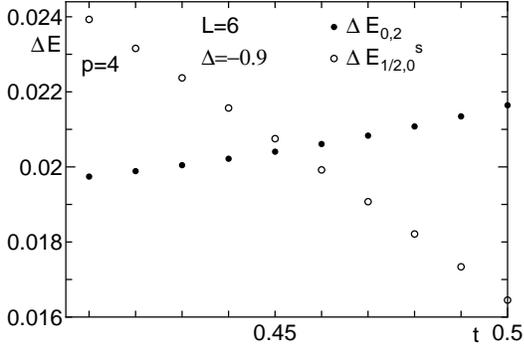}}
\caption{$j$-dependence of the energies $\Delta E_{0,2}$ and $\Delta E_{1/2,0}^{s}$ represented by $\bullet$ and $\circ$, respectively, for $p=4$, $\Delta=-0.9$ and $L=6$.}
\label{twie}
\end{figure}

\begin{figure}
\epsfxsize=70mm 
\centerline{\epsfbox{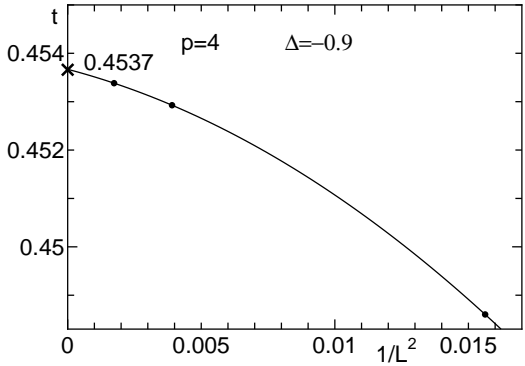}}
\caption{The extrapolation of $t_{c}$ for $p=4, \Delta=-0.9$.}
\label{ext}
\end{figure}

Figure \ref{twie} shows the behavior of $\Delta E_{0,2}$ and $\Delta E_{1/2,0}^{s}$ for $L=6$, $\Delta=-0.9$ and $p=4$. From the crossing point, we obtain $t_{c}(L=6)=0.4534$. The BKT transition point for the infinite system can be obtained by extrapolating from $L=2, 4$ and 6 to $L\rightarrow \infty$ as $t_{c}=0.4537$ assuming the extrapolation formula $t_{c}(L)=t_{c}+\frac{c_{1}}{L^{2}}+\frac{c_{2}}{L^{4}}$ as shown in Fig. \ref{ext} for $p=4$ and $\Delta=-0.9$. The extrapolated value is represented by $\times$.

\begin{figure}
\epsfxsize=70mm 
\centerline{\epsfbox{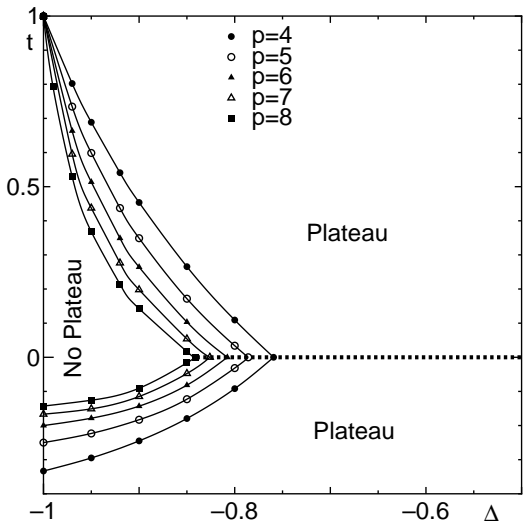}}
\caption{Phase boundaries on the $\Delta-t$-plane for $p=4,5,6,7$ and 8. The solid lines are the BKT plateau-non-plateau critical lines and the dotted line is a Gaussian plateau-plateau critical line. The lines are visual aids.} 
\label{phase}
\end{figure}

Using the same method, we determine the phase boundaries for $p=4,5,6,7$ and 8. The phase diagrams are shown in Fig. \ref{phase}. There are two BKT transition lines which merge on the multicritical point $\Delta=\Delta_c$ and $t=0$. The line $t=0, \Delta > \Delta_c$ is Gaussian corresponding to the uniform XXZ chain which has no plateau for any value of magnetization. 

From the Fig. \ref{phase}, we find that the non-plateau region becomes narrow and  the multicritical point $\Delta_{c}$ appraoches $-1$ with increasing periodicity $p$. These numerical results of multicritical points can be checked by the Bethe ansatz method as explained in the next section.

\section{Multicritical points determined by Bethe ansatz method}
By the numerical methods, the $p$-dependence of the multicritical points can not be calculated for large values of $p$. However, the multicritical point is located on the line $t=0$ which corresponds to a uniform XXZ chain. We can therefore make use of the Bethe ansatz solution which is available for arbitrary values of magnetization.\cite{yy,dc}. 

Within the XY-like region $-1<\Delta<1$, $\Delta$ is parametrized as $\Delta=\cos \theta$. The magnetization $m^{z}$ and Luttinger liquid parameter $K$ are determined in the following way. First, introduce the function $\sigma(\eta)$ for the density of particles satisfying the equations,
\begin{equation}
\label{sigma}
\sigma(\eta)=\frac{1}{2\pi} \Big \{ g(\eta)-\int_{-\Lambda}^{\Lambda} A(\eta-\eta')\sigma(\eta'){\rm d}\eta' \Big \},
\end{equation}
\begin{equation}
\label{msigma}
\int_{-\Lambda}^{\Lambda}\sigma(\eta){\rm d}\eta=\frac{1}{2}(1-2m^{z}),
\end{equation}
where the kernel $A(\eta)$ and the inhomogeneous term $g(\eta)$ are given by
\begin{equation}
\label{kernel}
A(\eta)=\frac{\tan \theta}{\tan^{2} \theta \cosh^{2}\frac{\eta}{2}+\sinh^{2}\frac{\eta}{2}},
\end{equation}
\begin{equation}
\label{g}
g(\eta)=\frac{\cot \frac{\theta}{2}}{\cosh^{2}\frac{\eta}{2}+\cot^{2}\frac{\theta}{2}\sinh^{2}\frac{\eta}{2}}.
\end{equation}
The real parameter $\Lambda \ge 0$ in  eq. (\ref{sigma}) and (\ref{msigma}) describes the values of the spectral parameter $\eta$ at the Fermi surface. Solving eq. (\ref{sigma}), the magnetization $m^{z}$ can be calculated using (\ref{msigma}) for a given value of $\Lambda$. 

On the other hand, the Luttinger liquid parameter $K$ is determined by the solution of the integral equation for the dressed charge function  $\xi(\eta)$.
\begin{equation}
\label{xi}
\xi(\eta)=1-\frac{1}{2\pi}\int_{-\Lambda}^{\Lambda}A(\eta-\eta')\xi(\eta'){\rm d}\eta',
\end{equation}
as
\begin{equation}
\label{r}
K=2\xi^{2}(\Lambda),
\end{equation}

\begin{figure}
\epsfxsize=70mm 
\centerline{\epsfbox{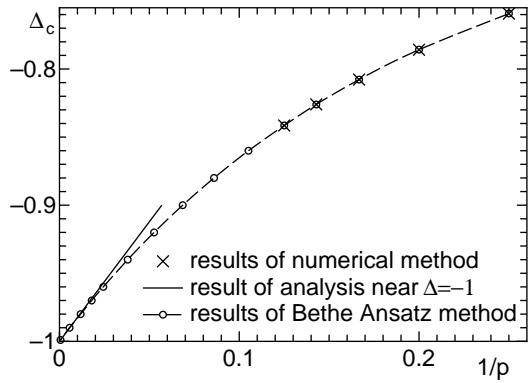}}
\caption{The $1/p$-dependence of the multicritical critical point $\Delta_{c}$. The results of the Bethe ansatz and the numerical method are represented by $\circ$ and $\times$, respectively. The solid line is the analytical result for $\Delta \rightarrow -1$. The dashed line is a visual aid.}
\label{vpm}
\end{figure}

At the multicritical point, the Luttinger liquid parameter $K$ is $4$. For the highest plateaus, the magnetization $m^{z}$ is $1/2-1/p$. Thus, the anisotropy parameter at the multicritical point $\Delta_{c}$ can be determined from  eqs. (\ref{sigma}), (\ref{msigma}), (\ref{xi}) and (\ref{r}). The $1/p$-dependence of the multicritical points $\Delta_{c}$ are shown in Fig. \ref{vpm}. The results of the Bethe ansatz method and the numerical method are represented by $\circ$ and $\times$, respectively. The results of numerical calculation coincide well with the Bethe ansatz results for $4 \leq p \leq 8$. 

Investigation of the behavior of $\Delta_c$ for large $p$ is now done in detail. Near $\Delta \simeq -1$, define an infinitesimal deviation of $\Delta$ from $-1$ by $\delta\Delta\equiv\Delta+1=\cos\theta+1\simeq (\pi-\theta)^{2}/2$ and $\delta\theta=\pi-\theta=\sqrt{2\delta\Delta}$. Since $\Lambda$ is small,  $\eta$ and $\eta'$ are also small. Therefore, eqs. (\ref{sigma}) and (\ref{xi}) can be transformed as
\begin{equation}
\label{sgx}
\sigma(x) = \frac{\delta\theta}{4\pi} + \frac{1}{2\pi}\int_{-\mu}^{\mu} dx' \frac{4\sigma(x')}{4 + (x-x')^2},
\end{equation}
\begin{equation}
\label{xix}
\xi(x) = 1 + \frac{1}{2\pi}\int_{-\mu}^{\mu} dx' \frac{4\xi(x')}{4 + (x-x')^2},
\end{equation}
where $x$ and $\mu$ are defined by $x=\eta/\delta\theta$ and $\mu=\Lambda/\delta\theta$. Comparing the inhomogeneous terms of  eqs. (\ref{sgx}) and (\ref{xix}), we get $\sigma(x)=\frac{\delta\theta}{4\pi}\xi(x)$. From eq. (\ref{msigma}), we find
\begin{eqnarray}
\label{magx}
1-2m^{z} &=& 2\delta\theta\int_{-\mu}^{\mu} dx \sigma(x) = \frac{\delta\theta^2}{2\pi}\int_{-\mu}^{\mu} dx \xi(x)\nonumber\\
&=&\frac{1+\Delta}{\pi}\int_{-\mu}^{\mu} dx \xi(x).
\end{eqnarray}
By solving eq. (\ref{xix}) numerically and adjusting the value of $\mu$ to satisfy $K=4$, we can determine the value of $\mu$ at the multicritical point. Using the thus obtained $\mu$ and $\xi(x)$, the relation between $\Delta_c$ and $m^{z}$ can be calculated by (\ref{magx}) as $1-2m^{z}=1.138352(1+\Delta_{c})$ ($m^{z}=1/2-1/p$). The resulting relation between $\Delta_c$ and $p$ is plotted in Fig. \ref{vpm} as a solid line. This result shows that the non-plateau region becomes narrow with increasing $p$ but does not vanish for finite $p$. It might appear strange that the  plateau region is enhanced in the large $p$ limit, because our model tends to the uniform XXZ chain in this limit. This is not surprising, however, because the magnetization at the plateau $m^z=1/2-1/p$ tends to the saturation magnetization $m^z=1/2$ in the large $p$ limit. The highest plateau state just tends to the fully magnetized state in this limit.

\section{Summary and Discussion}
The highest magnetization plateau state at $m^{z}=1/2-1/p$ in  $S=1/2$ $p$-merized  XXZ chain is investigated by exact diagonalization of finite size systems and the Bethe ansatz method for general values of $p$.  The BKT transition points are determined precisely by the level spectroscopy method with twisted boundary condition.\cite{ak,kiok,chs,no3} In the phase diagram, we find two plateau-non-plateau BKT transition lines, one plateau-plateau Gaussian line and one multicritical point for each $p$.  The non-plateau region becomes narrower with increasing periodicity. 

The multicritical points are calculated by the Bethe ansatz method and the results coincide well with the numerical results for $4\leq p\leq 8$. For large $p$, it is explicitly shown that the distance between the multicritical point $\Delta_c$ and the ferromagnetic point $\Delta=-1$ is proportional to $1/p$. This implies that the non-plateaus region does not vanish for finite $p$. As the periodicity $p$ becomes longer, the highest plateau tends to the fully magnetized state. 

 The tendency that the plateau region is enhanced with the periodicity $p$ is also found in the antiferromagnetic-(ferromagnetic)$_{n}$ polymerized $S=1/2$ chains\cite{chs}. From these examples, we may speculate that the plateau is stabilized as the size of the unit cell increases in general. It would be interesting to investigate a wider variety of models with long spatial periodicities to confirm this speculation. 

In this work, we confined ourselves to the $S=1/2$ $p$-merized XXZ chains with identical strength of modulation $t$ for the $XY$ and Ising components of the exchange coupling. If mapped onto the spinless fermion chains, the former corresponds to the transfer modulation and the latter, the nearest-neighbour interaction modulation. These two are, in principle, independent quantities. In fermionic language, the plateau-non-plateau transition corresponds to the metal-insulator transition. The non-plateau state corrsponds to the state in which the band gap is destroyed by the attractive interaction.  In this context, it must be important to investigate the effect of transfer modulation and interaction modulation separately. This is left for future studies.

The numerical calculation was performed using the program package TITPACK version 2 coded by H. Nishimori on HITAC S820 and SR2201 at the Information Processing Center of Saitama University and HITAC SR8000 at the Supercomputer Center, Institute for Solid State Physics, the University of Tokyo. This work is supported by a research grant from the Natural Science and Engineering Research Council of Canada (NSERC) and a Grant-in-Aid for Scientific Research from the Ministry of Education, Science, Sports and Culture of Japan.

\end{document}